# Deep learning approaches to surgical video segmentation and object detection: A Scoping Review

**Author List:** Devanish N. Kamtam[1],* Joseph B. Shrager[1,2],* Satya Deepya Malla[1], Nicole Lin[1], Juan J. Cardona[3], Jake J. Kim[1], Clarence Hu[4]

* co-first authors

**Affiliations:**

1. Division of Thoracic Surgery, Department of Cardiothoracic Surgery, Stanford University School of Medicine, Stanford, California, USA.
2. Veterans Affairs Palo Alto Health Care System, Palo Alto, CA, USA.
3. Department of Neurosurgery, Stanford University School of Medicine, Stanford, California, USA.
4. Hotpot.ai, Palo Alto, California, USA.

**Corresponding author:**

Devanish N. Kamtam

Division of Thoracic Surgery,

Department of Cardiothoracic Surgery

Stanford University School of Medicine

300 Pasteur Drive, MC 5407

Stanford, CA 94305-5407

Email: devanish@stanford.edu

**Abstract**:

**Introduction:**
Computer vision (CV) has had a transformative impact in biomedical fields such as radiology, dermatology, and pathology. Its real-world adoption in surgical applications, however, remains limited. We review the current state-of-the-art performance of deep learning (DL)-based CV models for segmentation and object detection of anatomical structures in videos obtained during surgical procedures.

**Methods:**
We conducted a scoping review of studies on semantic segmentation and object detection of anatomical structures published between 2014 and 2024 from 3 major databases - PubMed, Embase, and IEEE Xplore. The primary objective was to evaluate the state-of-the-art performance of semantic segmentation in surgical videos. Secondary objectives included examining DL models, progress toward clinical applications, and the specific challenges with segmentation of organs/tissues in surgical videos.

**Results:**
We identified 58 relevant published studies. These focused predominantly on procedures from general surgery [20(34.4%)], colorectal surgery [9(15.5%)], and neurosurgery [8(13.8%)]. Cholecystectomy [14(24.1%)] and low anterior rectal resection [5(8.6%)] were the most common procedures addressed. Semantic segmentation [47(81%)] was the primary CV task. U-Net [14(24.1%)] and DeepLab [13(22.4%)] were the most widely used models. Larger organs such as the liver (Dice score: 0.88) had higher accuracy compared to smaller structures such as nerves (Dice score: 0.49). Models demonstrated real-time inference potential ranging from 5-298 frames-per-second (fps).

**Conclusion:**
This review highlights the significant progress made in DL-based semantic segmentation for surgical videos with real-time applicability, particularly for larger organs. Addressing challenges with smaller structures, data availability, and generalizability remains crucial for future advancements.

**Keywords**: Computer vision, deep learning, semantic segmentation, object detection, surgical videos, tissues, organs.

**Abstract word count**: 244

**Introduction**:

Over the past decade, computer vision (CV) has emerged as a transformative technology in biomedicine. The advancement of CV and its application to biomedical images has resulted in remarkable breakthroughs across domains such as radiology [1], dermatology [2], and pathology [3], sometimes surpassing even human expert-level performance. This has led to several FDA-approved applications in clinical practice [4]. However, CV is yet to make a meaningful impact in surgical applications, in spite of the rich image and video data generated through minimally invasive surgery. Leveraging these vast data may enable CV-based surgical scene understanding and numerous promising surgical applications. Broadly these include facilitation of real-time tasks in the operating room (OR), such as the identification of critical structures, decision-making support during complex surgical procedures, intraoperative navigation, surgical workflow optimization, and even perhaps autonomous robotic surgery [5], [6]. Other promising applications include post-operative tasks such as automated generation of surgical reports, surgical education for medical students and surgical trainees, and automated assessment of technical proficiency for surgeons.

Surgical scene understanding is typically studied using tasks including, but not limited to, phase recognition, step recognition, image/video classification, object detection, semantic segmentation, and instance segmentation. Notably, object detection and semantic segmentation have been studied to identify various objects and their locations in a surgical scene. Object detection involves generating a bounding box for the detected object class, whereas the more challenging task of semantic segmentation requires precise pixel-wise classification of the image/video into any of the visible object classes in a surgical scene. Broadly, the objects to be identified or segmented in a surgical scene include surgical tools [7], which are relatively easier to detect; and organs/tissues, which are more challenging given their dynamic deformability [8], [9].

Understanding surgical scenes with CV has remained challenging due to the dynamic nature of surgical videos, which is often compounded by significant background noise, unlike more static and relatively less variable radiological and pathological images. Challenges with surgical videos include the lack of large, labeled video datasets and the substantial costs of obtaining high-quality annotations for supervised learning. This problem is further compounded by the frequency of human anatomical variations and differences in individual surgeons' surgical approaches/techniques. Furthermore, continuous alterations in tissue appearance as a result of real-time surgical actions such as exposing, retracting, cutting, dissecting, and cauterizing tissues -- and interferences from fogging, varying light intensity, shadows, surgical smoke, blood, and fluid stains -- create substantial background noise, posing significant challenges for CV models to analyze and interpret these videos effectively [10].

Deep learning (DL) has proven particularly effective in addressing these challenges, offering substantial advancements over traditional image analysis and machine learning approaches such as change detection algorithm [11], Fourier transform [12], support vector machines [13], [14], phase change algorithm [15], and random forests with pulsation analysis [16]. DL-based methods of image analysis, most notably convolutional neural networks (CNNs) and their modifications, such as U-Net, DeepLab, and Mask R-CNN, and most recently transformer-based architectures, have enabled understanding of surgical scenes through tasks like object detection and semantic segmentation. Given the rapid advancement of DL over the past few years, there is a need to review the current landscape of DL in surgical scene understanding for these tasks.

Hence, this scoping review aims to summarize the literature published to date on the evaluation of state-of-the-art DL models for segmentation and object detection of anatomical tissues in surgical procedures. It also highlights the challenges faced, and it identifies research gaps and future directions for advancing CV applications in the surgical domain.

### 1.1 Contributions

The key contributions of this review include:

- We conduct the first scoping review of studies on the semantic segmentation and object detection of anatomical structures in real-world videos across all surgical procedures.
- We present the current state-of-the-art for semantic segmentation of various organs or tissues in surgical videos or images.
- We identify the DL models employed for semantic segmentation and object detection in the selected studies, summarizing their performance and real-time applicability.
- We provide an overview of the surgical procedures investigated and the specific applications of segmenting various organs within these procedures.
- We highlight key challenges associated with semantic segmentation and object detection of anatomical structures across different surgical procedures, summarize solutions proposed in the studies to address these challenges, and outline the potential directions for future research.

## 2. Research Methodology

This scoping review follows the PRISMA-ScR (Preferred Reporting Items for Systematic Reviews and Meta-Analyses-Scoping Review) [17] framework and reporting guidelines to ensure methodological rigor and reproducibility. All reporting aligns with the PRISMA-ScR checklist. The study protocol was not registered with PROSPERO, as several of the included studies focus on technical aspects, which lie beyond the scope of PROSPERO's primary emphasis on clinical outcomes and patient-centered research.

### 2.1 Research Questions

We identified a primary research question to guide this scoping review:

- What is the current state-of-the-art performance for semantic segmentation of various organs/tissues using real-world surgical videos?

Secondary objectives, designed to gain a comprehensive understanding of this topic, were also established:

- What DL models and model architectures have been applied for semantic segmentation and object detection of anatomical structures in surgical videos?
- For which surgical procedures have DL been used for semantic segmentation and object detection of anatomical structures in surgical videos?
- What were the clinical applications of computer vision tasks, such as semantic segmentation and object detection, across various surgical procedures?
- What were the major challenges associated with applying DL-based semantic segmentation and object detection methods for anatomical structures in surgical videos?

These research questions informed the search strategy, study inclusion criteria, and evaluation framework for the selected studies.

### 2.2 Search Strategy

We queried the following literature databases for studies published from January 2014 to September 2024, since the field began emerging after 2012 and our focus was on recent approaches.

· PubMed (https://pubmed.ncbi.nlm.nih.gov/)

· Embase (https://www.embase.com/)

· IEEE Explore (https://ieeexplore.ieee.org/Xplore/home.jsp)

These databases were selected to ensure a comprehensive coverage of publications spanning both technical and clinical domains. The search was conducted using the following search strings and Boolean logic.

(Instance Segmentation OR Semantic Segmentation OR Object Detection OR Object Recognition OR Object Localization) AND (Anatomical OR Tissues OR Organs OR Nerves OR Vessels OR Arteries OR Veins OR Lymph Nodes) AND (Surgery OR Surgical OR Operative OR Intraoperative OR Laparoscopic OR Robotic) AND (Deep Learning OR Artificial Intelligence OR Neural Network OR Computer Vision)

MedRxiv (https://www.medrxiv.org/) and arXiv (https://arxiv.org/) were not searched due to the absence of an advanced search query mechanism.

### 2.3 Study Selection Criteria

The study selection process was conducted in two stages. In the first stage, titles and abstracts of the search results obtained using the predefined search criteria were screened to identify potentially relevant studies. The second stage involved a detailed full-text review to create a final list of studies relevant to the subjects of the review. Additional studies were identified through snowballing and recursive manual searches of the reference sections, which were subsequently subjected to quality assessment and detailed evaluation of study findings. The screening and study selection process was conducted by a single reviewer (D.N.K. and J.J.K.) and disagreements were resolved through discussion.

The inclusion and exclusion criteria were as follows:

- **Inclusion criteria**:
    - Semantic segmentation and object detection of anatomical tissues/organs as the primary task.
    - Videos/images of live human subjects.
    - Published from 2014 onwards to stay relevant to recent/advanced technologies.
    - Peer-reviewed journal and conference manuscripts (PubMed, Embase, IEEE Explore), as well as non-peer-reviewed preprints (arXiv).
- **Exclusion criteria**:
    - Surveys and reviews.
    - Non-availability of full-text manuscripts.
    - Non-English studies.
    - Only instruments (i.e., no anatomical structures) as the object classes for segmentation/detection.
    - Non-human and cadaver-based studies.
    - Performance metrics not reported.

### 2.4 Data extraction

During the data extraction process, a predefined extraction form was used to systematically collect relevant information from the shortlisted studies. The extracted data is provided in the Supplementary Data File 1. Key variables collected included study characteristics such as the title, authors, year of publication, aim of the study, dataset characteristics (e.g., newly generated or previously existing data), dataset availability (public or private), evaluated tasks, dataset size (number of videos and annotated frames), video resolution, operative approach (laparoscopic/robotic/microscopic), surgical procedures evaluated, models employed, inference times, and the outcomes measured.

The extracted outcome metrics included the Jaccard index/Intersection over Union (IoU), Dice similarity coefficient (DSC)/F1 score, precision, recall, sensitivity, specificity, false positivity rate, false negative rate, accuracy, and area under precision-recall curve (AuPROC). Summary metrics such as mean average precision (mAP), mean IoU, and mean DSC were collected when reported specifically for anatomical tissues. If the summary metrics included both anatomical tissues and surgical instruments, the range of individual scores for the evaluated anatomical tissues was presented. In instances where individual tissue scores were unavailable, the overall summary metric encompassing both tissues and instruments was reported.

### 2.5 Study Quality Assessment

The quality assessment of the included studies was conducted using the CLAIM (Checklist for Artificial Intelligence in Medical Imaging) 2024 [18], as it was considered relevant for this review's focus on the application of AI to medical images. However, given the unique characteristics of surgical videos/images compared to other medical or radiology images, we selectively applied the criteria pertinent to surgical videos. The variables used to assess study quality included:

- Impact factor of the published journal (in the year of publication)
- Description of clinical demographic data of included videos
- Patient-level split of the included video data/frames
- Frame sampling methodology (random or selective)
- Annotation source (board-certified surgeon or non-medical annotator)
- Reporting of inter-rater variability
- Ablation studies performed
- Benchmarking against other models
- Evaluation on internal testing data (validation or test)
- Evaluation on external testing data
- Qualitative failure analysis performed
- Availability of code (for reproducibility)
- Availability of video dataset and annotations

The data extracted for this quality assessment is provided in Supplementary Data File 2, which includes detailed information for each evaluated study. This assessment was conducted independently by two reviewers (J.J.K. and D.N.K.) and disagreements were resolved through discussion.

No studies were excluded based on this quality assessment to ensure the availability of an adequate number of studies for a comprehensive review. This decision was also guided by the absence of a widely accepted quality assessment framework specific to AI-based evaluation of surgical videos.

**Results**:

We screened 1141 search results using the predefined search criteria and identified 142 studies as potentially relevant. We then narrowed them down to 46 studies after a detailed review of the titles, abstracts, and full manuscripts. In addition, 12 studies were included after manually searching through the references of the included studies, resulting in a total of 58 studies. Of these, 54 (93.1%) were peer-reviewed. All the selected studies were published in or after 2019. The PRISMA flowchart is presented in Figure 1, and the annual publication trend is charted in Figure 2. The list of all studies, year of publication, first authors, surgical procedures investigated, and target organ classes are provided in Supplementary Table 1.

Of the 58 studies reviewed, 13 (22.4%) utilized pre-existing segmentation datasets, whereas 45 (77.6%) created their own segmentation datasets by annotating masks using their institutional video collections or other publicly available non-segmentation datasets. Of these newly created datasets, 8 (17.7%) were made publicly available, whereas 37 (82.2%) remained private. The median dataset size was 42.5 videos (IQR, 17-100.5), with the largest dataset comprising 548 videos. The median frames/images count was 1,963 (IQR, 672.5–4,053.7) prior to dataset augmentation, while the largest dataset contained 47,241 frames. Data augmentation was employed by 32 studies (55.2%). Regarding video quality, 17 studies (29.3%) used datasets with a resolution of ≥1080p, while 30 studies (51.7%) used datasets with ≥720p resolution.

**Semantic segmentation performance metrics for various organs**

The segmentation performance within each study by various metrics of segmentation accuracy, such as DSC, mIoU, etc., is presented in Supplementary Table 2. Organs with the highest median DSCs included the liver (0.88), lung (0.89), spleen (0.85), and kidney (0.86). In contrast, certain tissues had notably lower median DSCs: intestinal veins (0.49), nerves, superior hypogastric plexus (0.49), Glissonean pedicles (0.48), vesicular glands (0.43), tegmen (0.29), and omentum (0.18). The median DSC and mean IoU of each organ class across all studies in the dataset are summarized in Supplementary Table 3. And the state-of-the-art organ-specific DSC for each study is provided in Supplementary Data File 3.

**Model characteristics and inference time**:

Of the 58 studies reviewed, 31 (53.4%) used established models, while 27 (46.5%) introduced novel customized technical modifications to existing models. The most commonly employed models were non-transformer-based models in 46 (79.3%) studies. Among these, U-Net-based models were the most common [14 (24.1%)], followed by DeepLab-based models [13 (22.4%)]. Transformer-based models were employed in around 7 (12.1%) studies (Table 1).

Most models demonstrated real-time inference potential, with the highest reported speeds as follows: 298 frames per second (fps) using TensorRT with FP16 precision [19], 233 fps with FP32 precision [19], 141.5 fps with U-Net [20], 137 fps with U-net [21], and 87.44 fps with YOLACTEdge optimization [22].

**Surgical procedure characteristics**:

The procedures investigated were from several specialties, including general surgery (20/58, 34.4%), colorectal surgery (9/58, 15.5%), and neurosurgery (8/58, 13.8%). The most common procedures included cholecystectomy (14/58, 24.1%), anterior rectal resection (5/58, 8.6%), cataract surgery (4/58, 6.9%), and liver resection, prostatectomy and nephrectomy (each 3/58, 5.1%). There were only 2 procedures, esophagectomy (2/58, 3.4%) and lung resection (2/58, 3.4%) investigated from the field of thoracic surgery – the particular interest of the authors of this review. The most common approaches for video data collection were laparoscopic (34/58, 56.8%), microscope-assisted (10/58, 17.2%), and robot-assisted (9/58, 15.5%) (Table 2).

**Computer vision tasks and applications**:

The tasks investigated included semantic segmentation (47/58, 81.0%), object detection (14/58, 24.1%), landmark detection (6/58, 10.3%), and instance segmentation (2/58, 3.4%). The clinical applications were primarily directed at real-time intraoperative guidance in identifying anatomical structures in an attempt to enhance surgical safety and precision. These applications included, specifically:

· identifying structures within the hepatocystic triangle to confirm the critical view of safety before clipping and cutting the cystic duct, thereby attempting to reducing the risk of common bile duct injury

· identifying the adrenal vein and renal artery during renal and adrenal surgeries to putatively enable more precise dissection and prevent catastrophic bleeding

· identifying the prostate during transanal total mesorectal excision (TaTME) to attempt to avoid urethral injuries

· identifying loose connective tissue to define safe dissection planes during gastrectomy, attempting to reduce the risk of a postoperative pancreatic fistula

· identifying the recurrent laryngeal nerve during thyroidectomy and esophagectomy to attempt to reduce the incidence of nerve injuries/hoarseness/swallowing dysfunction

· identifying the thoracic nerves during lung resections to attempt to avoid injury to these

· identifying ocular structures during cataract surgery to attempt to facilitate autonomous cataract surgery

· identifying autonomic nerves, ureter, and inferior mesenteric vessels during colorectal surgery to attempt to avoid injury to these structures

· identifying the vena cava and azygous vein during esophagectomy to attempt to prevent vascular injury

· identifying the ureter during hysterectomy to attempt to reduce the risk of ureteral injury

- identifying parathyroid glands during thyroid surgery to attempt to avoid postoperative hypocalcemia

**Discussion**:

Our scoping review on CV identified 58 studies investigating the use of DL for live organ/tissue segmentation in surgical videos/images. Among these, 45 studies created new annotated datasets, while 13 utilized existing ones. Notably, only 17.7% were made publicly available. The surgical specialties predominantly studied were general surgery, colorectal surgery, and neurosurgery. And the most frequently studied procedures were cholecystectomy and anterior resection of the rectum. Most studies were directed at intraoperative guidance in identifying critical structures to reduce the risk of injuries during surgeries. There were only four studies that addressed Thoracic Surgery. While the earlier studies predominantly employed variants of traditional CNNs, particularly U-Net and DeepLab versions, recent years saw a growing adoption of transformers and other attention-based models.

### Segmentation performance, challenges, and real-time inference

CV through DL-based segmentation performed comparably to specialist surgeons in detecting nerves [23], outperformed all but the most experienced surgeons in segmenting pancreas [24], and proved to be a more effective educational tool for trainees versus traditional surgical learning [25]. Notably, segmentation of tissues improved performance on clinical applications such as, determining critical view of safety in the complex anatomy of porta hepatis [26]. Thereby, CV can enable intraoperative navigation for various surgeries without the need for additional hybrid OR equipment [27]. Specifically, large solid organs with consistent textures or high local similarity, such as the liver, lung, and spleen, had the highest Dice scores. Whereas the smaller, sparsely distributed tissues, such as nerves and veins, had lower scores. Additional challenges were encountered at tissue boundaries partially obscured by fat or fascia [21], [28], which was also challenging for expert surgeons' annotation, resulting in only moderate to substantial interrater agreement (IRA) (0.58 -0.86) [29], [30], [31]. Notably, studies that binarized organs into exposed and partially exposed organs resulted in better Dice scores by ~3%.

However, given that surgeons may perceive organ boundaries as probabilistic rather than binary [32], more advanced methods of annotating or modeling organs may be needed for CV to better understand surgical videos. Inference time is another key consideration for real-time application of CV in surgery. While inference at ≥30 fps (or ~30 ms/frame) is widely considered as necessary, 11 fps has also been considered acceptable [28]. Accordingly, several models such as U-Net, YOLO, YOLACT, and DeepLabv3 have exhibited these speeds without any discernible lag [20], [33]. However, given the variations in image resolutions and graphics processing units (GPUs) used, identifying the models with the best inference times was challenging. And as faster inference times often come with trade-offs in accuracy [34], future research could focus on the optimal combination of both for realizing real-world utility.

### Models and model architectures

Segmentation models must account for local context that is often disrupted by noise, such as blood, smoke, etc., and global context that is influenced by tissue deformability. Most state-of-the-art semantic segmentation networks are based on fully convolutional neural networks (FCNs). U-Net is the most widely used FCN for biomedical segmentation. It captures local granular features and the local spatial context very effectively. Larger contexts can be captured by U-net by broadening/enhancing the receptive fields, but this comes at the expense of features of finer resolution. While U-net overcomes this partly by preserving the lower-level features through skip connections, it may still struggle to capture details such as fine blood vessels [21]. Therefore, despite the success of CNNs in biomedical segmentation, they are constrained by these limited receptive fields that reduce their ability to effectively capture and learn global features [35].

Especially in surgery, beyond the local context/fields captured by traditional CNN architectures, inferring global or long-term dependencies is also crucial. Various studies have addressed this, using techniques such as pyramid networks with atrous convolutions, pyramid pooling, multi-scale feature fusion, attention-based, and transformer architectures. Among these, atrous convolution incorporated by DeepLabv3, a U-Net-based model, showed improved retention of global information [36], [37], but occasionally struggled to preserve finer details, such as edges [9]. To overcome this, Deeplab V3+, combined atrous spatial pyramid pooling with atrous convolution to achieve fusion of multi-scale networks and optimize edge accuracy [38]. Notably, U-Net++ was also able to acquire global features by redesigning skip pathways and dense connections enabling a smoother merging of the global and local features [9]. Lastly, hybrid architectures combining transformers with CNNs, such as UNETR [36] and TransFuse [37], have also been introduced with varying performance improvements. However, the significant variability in procedures and organs evaluated across studies made it difficult to determine the most effective model.

### Data diversity, data augmentation, and unsupervised approaches

While models are pivotal for achieving high segmentation performance, considerations about training data are equally crucial. There remains a critical need for large, labeled datasets to advance surgical vision models for real-world applications. Equally important is ensuring data diversity, as class imbalance—whether in the frequency of organ appearances within scenes or size/pixel representation—poses significant challenges. This has been tackled by repeat factor sampling and adaptive sampling, which have significantly boosted performance on rare classes, regardless of model architecture [39]. Yet, the primary challenge in acquiring training data is the limited availability, high costs of expert annotators, i.e. surgeons, and variability in IRA that is also observed among expert surgeons [29], [30], [31]. Proposed solutions to minimize variability include integrating masks from multiple annotators or implementing a two-stage annotation process, where non-surgeons perform the initial annotations, followed by expert surgeons' revisions. Importantly, while data augmentation can enhance training efficiency, improper augmentation can negatively impact model performance [40] – e.g., horizontal or vertical flips in training data that do not naturally occur can misguide the model's weights. Lastly, unsupervised approaches also offer effective solutions to address the lack of labelled training data. Some unsupervised approaches explored include dataset reconstruction through pretraining [35], local semantic consistency for mask generation [41], auxiliary image reconstruction [42], incorporating ICG-fluorescence with Otsu-thresholding [10], and hyperspectral imaging information [43].

### Evaluation metrics and loss function

The choice of metrics for optimizing segmentation performance is also a key consideration. Metrics such as specificity or accuracy are heavily influenced by true negatives and can produce misleadingly high results [44]. In contrast, metrics like DSC or IoU that penalize false positives offer a more accurate assessment of model performance. Regarding loss function, Dice loss was particularly effective for imbalanced classes compared to the cross-entropy loss [8], which demonstrated poor correlation with the actual performance metric, i.e., mIoU [39].

Despite its advantages, Dice loss is not without limitations, as its lack of differentiability can hinder optimization. Direct training with the mIoU metric is also hindered by non-differentiability. Consequently, surrogates like the Lovasz-Softmax extension of the Jaccard index/IoU have been explored and shown to optimize the IoU metric more effectively [39]. Other approaches have also investigated online hard example mining, which ignores loss for pixels where the correct label is predicted with a probability >0.7, to focus training on more challenging examples.

### Foundational pretrained models and generalizability

Variations in surgical approaches, even for similar procedures — such as transperitoneal or retroperitoneal for nephrectomy; and McKeown, Ivor-Lewis, or transhiatal approaches to esophagectomy — create diverse data needs, necessitating large volumes of labeled data to train task-specific models. By contrast, pretrained foundational models such as large language models, have shown significant utility in other domains by reducing the need for extensive task-specific data through transfer learning. Similarly in segmentation, pre-trained features from models trained on ImageNet/COCO have shown superior segmentation compared to task-specific models trained from scratch [20], [22], [29]. Pretraining not only enhances performance and reduces training data needs but also mitigates overfitting and improves model generalizability [45]. This generalizability potential in surgical segmentation is exemplified by a model that was trained to identify nerves in colorectal surgeries, yet accurately detected nerves in gastrectomy procedures [25]. The success of recently introduced foundational segmentation models, such as Segment Anything Model (SAM) [46] and SAM 2 [47], across several domains [48] holds immense potential to further advance surgical segmentation performance by leveraging their generalized representations.

### Current real-world applications

Despite significant advances in surgical video segmentation and object detection by DL models, the clinical applications of CV remain largely confined to the research phase.  These tasks mainly include identifying nerves, vessels, and organs for intraoperative navigation in an effort to enhance surgical safety and precision. CV through tissue segmentation holds promise in realizing real-time operative guidance, automated skill assessment, surgical workflow analysis, and augmented reality (AR)-based training and education with real-time feedback for surgical trainees. While the potential to reduce complications and improve surgical outcomes is clear, the adoption of these models in real-world ORs remains limited. This is primarily due to challenges in model generalizability, driven by the lack of large diverse datasets. Bridging the gap between research innovations and clinical applications is essential in realizing the transformative potential of computer vision in surgery.

**Research Directions**:

- Developing large open source labeled diverse datasets: there is a need for comprehensive and diverse datasets with annotations for several tasks relevant to surgical scene understanding.
- Employing generalized foundational models: there is a need to employ foundational models such as SAM that generalize across most organs vs. organ-specific models to reduce the requirement for massive datasets.
- Maintaining a focus upon unsupervised training approaches: focusing on unsupervised learning methods to reduce dependency on annotated data.
- Optimizing for real-time inference while preserving/enhancing the accuracy of segmentation.
- Investigating real-world applicability and utility of these models in the proposed applications to enable translation into clinical practice.

**Conclusion**:

In conclusion, this review highlights the advancements that have been made, and the challenges that remain, in applying deep learning in computer vision for semantic segmentation and object detection of tissues/organs in surgical videos. Deep learning models have achieved state-of-the-art segmentation performance for most anatomical structures with promising real-time capabilities. However, challenges remain, including the limited availability of large-scale, diverse datasets and the task-specific nature of these segmentation models. Addressing these gaps through larger, diverse datasets, and leveraging foundational segmentation models through unsupervised approaches is crucial to further improve semantic segmentation and object detection in surgical videos and realize the full potential of computer vision in surgical applications.

**CRediT authorship contribution statement:** DNK – Conceptualization, Data curation, Formal analysis, Methodology, Supervision, Writing – original draft, Writing – review and editing; SDM - Conceptualization, Data curation, Writing – original draft, Writing – review and editing; NL - Conceptualization, Methodology, Writing – review and editing; JJC – Conceptualization, Data curation, Methodology, Writing – review and editing; JJK - Data curation, Formal analysis, Methodology, Writing – original draft; CH - Conceptualization, Methodology, Supervision, Writing – review and editing; JBS - Conceptualization, Methodology, Supervision, Writing – review and editing.

**Declaration of generative AI and AI-assisted technologies:** During the preparation of this work, the author(s) utilized ChatGPT to assist with rephrasing and refining the writing in the manuscript. After using this tool/service, the author(s) reviewed and edited the content in detail and take(s) full responsibility for the content of the published article.

**Conflicts of interest**: Joseph Shrager: Consulting – Lungpacer, Inc.; Becton Dickinson, Inc.; Clarence Hu: Founder – Hotpot.ai; Other authors have nothing to declare.

**Funding**: None

**Data statement and data linking**: The data generated and analyzed during the current study are publicly available in Figshare (10.6084/m9.figshare.28418759).

**Ethical approval**: Not required for reviews

**Acknowledgements**: None

**Tables**:

**Table 1**: State-of-the-art deep learning models in all the studies stratified based on the major mechanism of their architecture.

| **Basic mechanism** | **Broad model** | **Models** |
|---|---|---|
| Non-transformer based CNNs (46) | Traditional CNNs (6) | CNNs (3) |
| | | FCN (3) |
| | U-net (14) | U-net (5) |
| | | U-net with other encoders (3) |
| | | Modified U-net (3) |
| | | U-net++ (1) |
| | | Mask-R-CNN (2) |
| | DeepLab (13) | DeepLabv3 (1) |
| | | DeepLabv3+ (8) |
| | | Modified DeepLabv3+ (1) |
| | | DeepLabv3+Resnet (2) |
| | | DeepLabv3+ with CNN (1) |
| | YOLO (6) | YOLO (3) |
| | | YOLOv3 (1) |
| | | YOLO with attention (1) |
| | | YOLOv4 (1) |
| | YOLACT (3) | YOLACT (1) |
| | | YOLACT++ (2) |
| | Other models (4) | Feature pyramid network (4) |
| Transformer based CNNs (7) | | Transformer-based (6) |
| | | Transformer plus CNN (1) |
| Other models (5) | | Custom models (4), SOLO (1) |

CNN, convolutional network; FCN, fully convolutional network.

**Table 2**: The distribution of surgical specialties, procedures, and segmented organs across all the studies included in the review

| **Surgical specialty** | **Surgical procedure** | **Organs segmented** |
|---|---|---|
| Colorectal surgery (10) | Anterior resection (6),<br><br>Sigmoid colectomy (3),<br><br>TaTME (2),<br><br>Right Colectomy (1) | autonomic nerves (hypogastric nerve and superior hypogastric plexus) (3), prostate (2), ureter (2), inferior mesenteric artery (2), small intestine (2), superior mesenteric vein (1), ileocolic artery (1), and ileocolic vein (1), superior rectal artery (1), inferior mesenteric vein (1), Areolar tissue in the TME plane (1), Abdominal wall (1), colon (1), liver (1), pancreas (1), spleen (1), stomach (1), vesicular glands, uterus (1) |
| ENT (1) | Simple mastoidectomy (1) | external auditory canal (1), antrum (1), tegmen (1), sigmoid sinus (1), spine of Henle (1) |
| General Surgery (20) | Cholecystectomy (15),<br><br>Hernia repair (2),<br><br>Thyroidectomy (2),<br><br>Adrenalectomy (1) | Gallbladder (12), liver (10), cystic duct (8), cystic artery (6), fat tissue (5), common bile duct (4), cystic plate (4), hepatocystic triangle (3), abdominal wall (3), gastrointestinal tract (3), left adrenal vein (1), connective tissue (1), hepatic vein (1), liver ligament (1), lower edge of the medial segment of liver (2), Rouviere's sulcus (2), pubic symphysis (1), direct hernia orifice (1), Cooper's ligament (1), Iliac vein (1), Triangle of Doom (1), deep inguinal ring (1), iliopsoas muscle (1), hepatic ligament (1), mesentery (1), colon (1), stomach (1), pancreas (1), omentum (1), parathyroid gland (1), uterus (1), recurrent laryngeal nerve (1) |
| Gynecology (2) | Hysterectomy (2) | uterus (1), ovaries (1), uterine artery (1), ureter (1), nerves (1) |
| Hepatobiliary-pancreatic surgery (3) | Liver resection (3),<br><br>Pancreaticoduodenectomy (1), Cholangiocarcinoma resection (1) | hepatic veins (1), Glissonean pedicles (1), IPDA (1), Henle trunk (1), PHA (1), CHA (1), GDA (1), RGEV (1), Portal vein (1), SMV (1), bile duct (1), bile duct transection site (1) |
| Neurosurgery (8) | Microvascular decompression (2), Pituitary resection (2),<br><br>Tumor resections (2), Spinal dural arteriovenous fistula surgery (1), Endoscopic third ventriculostomy (1) | cerebral vessels (arteries or veins) (3), trigeminal nerve (2), facial nerve (2), glossopharyngeal nerve (2), vagus nerve (2), anterior inferior cerebellar artery (2), posterior inferior cerebellar artery (2), petrosal vein (2), dorsal spinal arteries (1), spinal cord (1), Sella (1), clival recess segmentation (1), carotid artery (1), and optic protuberance (1), Choroid plexus (1), foramen of Monro (1), anterior septal vein (1), mammillary bodies (1), infundibular recess (1), dura mater (1), and tumor area (1) |
| Ophthalmology (5) | Cataract surgery (4), Vitreoretinal surgery (1) | pupil (4), iris (4), cornea (4), skin (4), optic disc (1), fovea (1), retinal tears (1), retinal detachment (1), epiretinal membrane (1), fibrovascular proliferation (1), endolaser spots (1), and global area where endolaser was applied (1) |
| Thoracic Surgery (4) | Esophagectomy (2), Lung resection (2) | azygos vein (1), vena cava (1), aorta (1), right lung (1), recurrent laryngeal nerve (1), thoracic nerves (1), Lung surfaces (1), tumor lesions on the lung surface (1) |
| Upper GI surgery (2) | Gastrectomy (2) | loose connective tissue fibers, pancreas (1), mesogastrium (1), transverse mesocolon (1), intestine (1), blood vessels (1), greater omentum (1), pre-pancreatic fatty tissue (1) |
| Urology (5) | Prostatectomy (3), Nephrectomy (3) | prostate (1), kidney (2), urinary bladder (1), seminal vesicle-vas deferens (2), renal artery (2), |

**Figures**:

**Figure 1**: The PRISMA flowchart

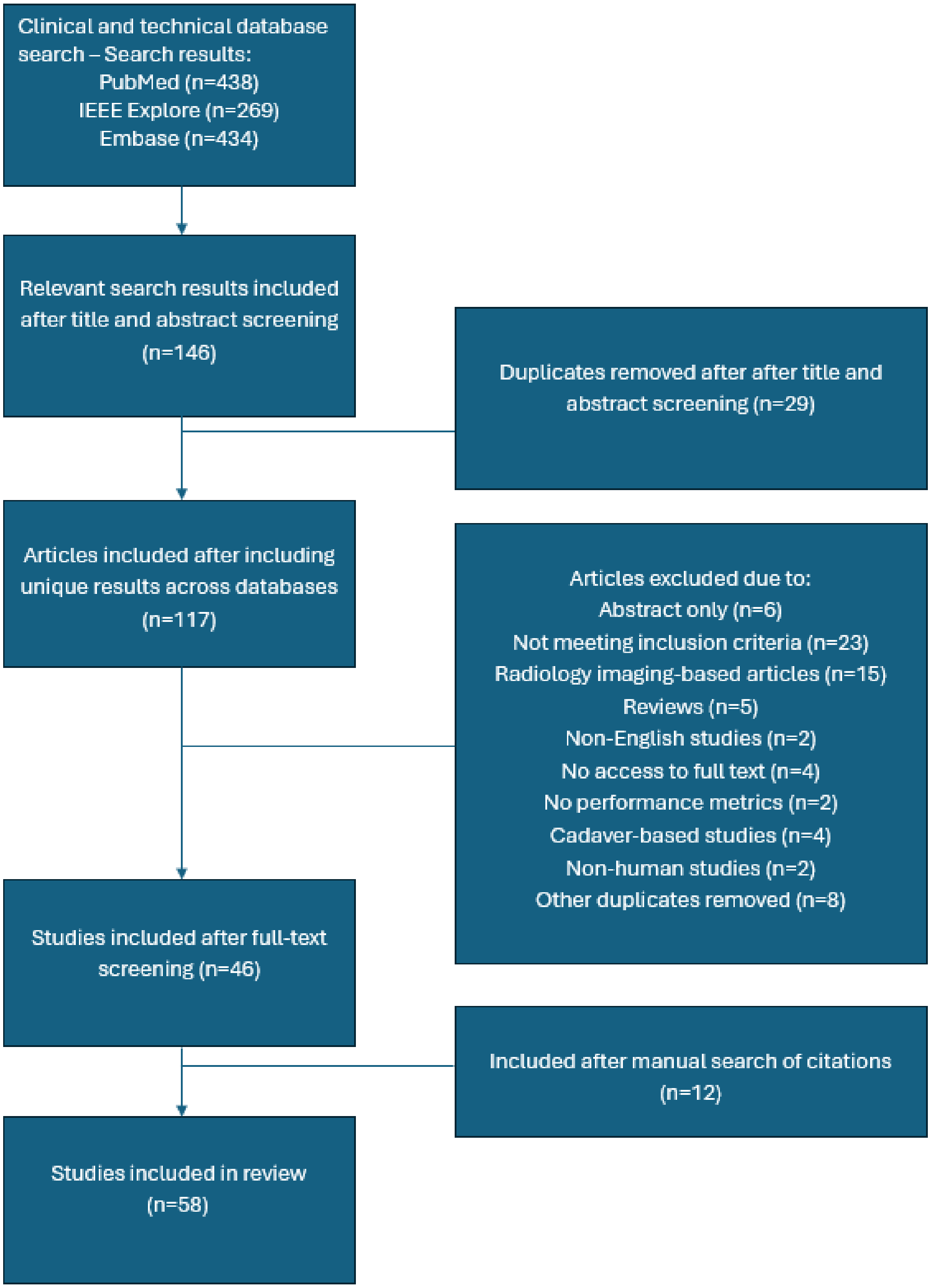

**Figure 2**: The distribution of studies across different years in the review from 2014 to 2024.

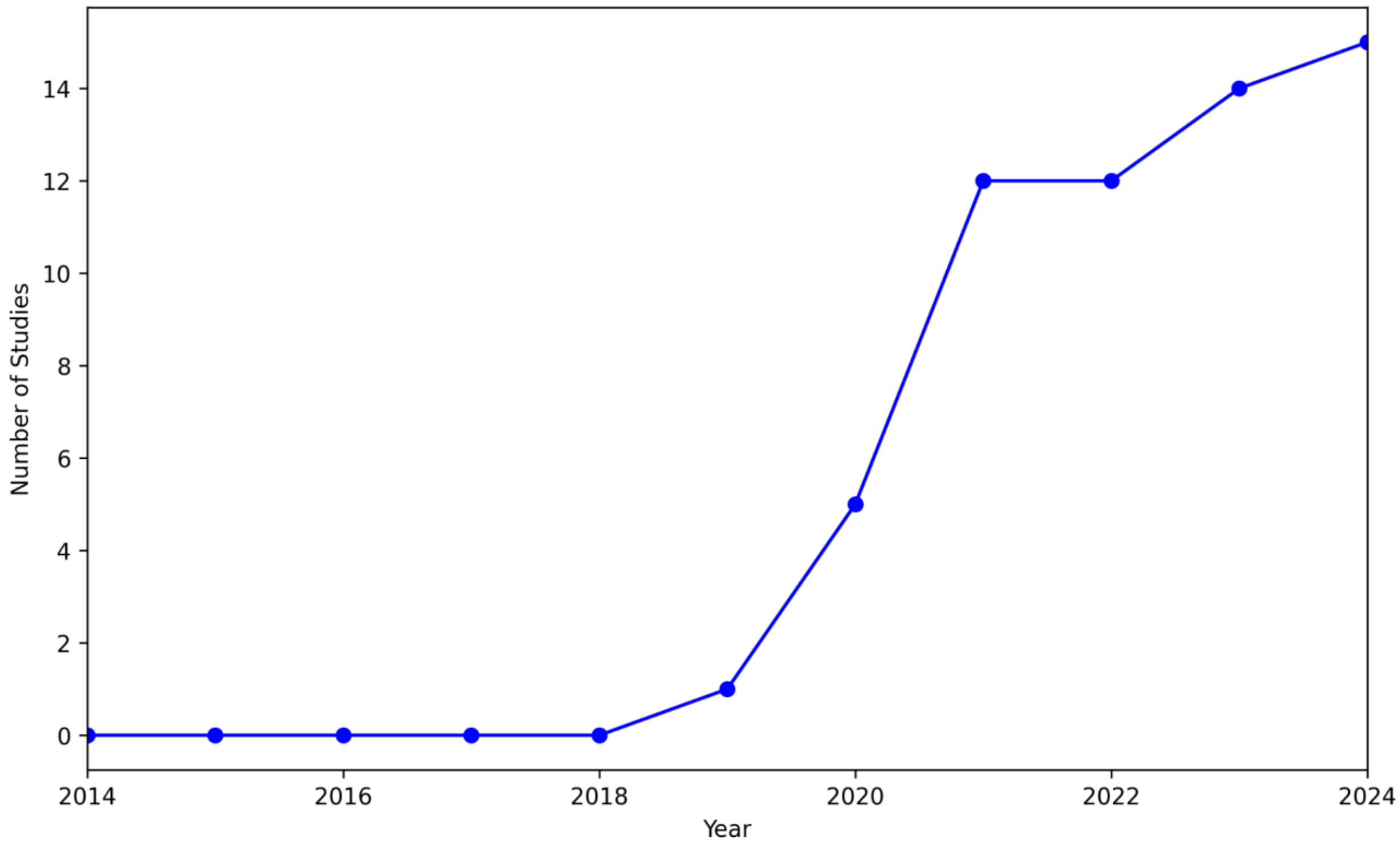

**Supplementary material**:

**Supplementary Table 1**: A description of the titles, years of publication, first authors, surgical procedures evaluated, surgical approach, and organs segmented in all the studies included in the review.

**Supplementary Table 2**: A description of data contained in all studies included in the review.

**Supplementary Table 3**: The segmentation metrics (dice similarity coefficient and mean intersection-over-union (mIoU)) of all the organs evaluated in the studies included in the review.

**Supplementary Data File 1:** A detailed description of the titles, years of publication, first authors, surgical procedures evaluated, surgical approach, and organs segmented in all the studies included in the review.

**Supplementary Data File 2:** A description of the detailed quality assessment data extracted from each evaluated study included in the review.

**Supplementary Data File 3:** The segmentation metrics (dice similarity coefficient and mean intersection-over-union (mIoU)) of all the organs evaluated in the studies included in the review.

**Supplementary Table 1**: A description of the titles, years of publication, first authors, surgical procedures evaluated, surgical approach, and organs segmented in all the studies included in the review.

| Title | Year | Authors | Surgical Procedure | Surgical Approach | Target organs |
|---|---|---|---|---|---|
| CaDIS: Cataract dataset for surgical RGB-image segmentation | 2019 | M Grammatikopoulou | Cataract surgery | Microscopic | Pupil, iris, cornea, skin |
| Artificial Intelligence for Intraoperative Guidance: Using Semantic Segmentation to Identify Surgical Anatomy During Laparoscopic Cholecystectomy | 2020 | A Madani | Cholecystectomy | Laparoscopic | Safe zone and dangerous zones of dissection, gallbladder, liver, hepatocystic Triangle |
| SurgAI: deep learning for computerized laparoscopic image understanding in gynaecology | 2020 | SM Zadeh | Hysterectomy | Laparoscopic | Uterus, ovaries |
| NephCNN: A deep-learning framework for vessel segmentation in nephrectomy laparoscopic videos | 2020 | Alessandro Casella | Partial nephrectomy | Robot | Renal arteries |
| Deep learning for semantic segmentation of organs and tissues in laparoscopic surgery | 2020 | Paul Maria Scheikl | Cholecystectomy | Laparoscopic | Liver, gall bladder, fat |
| m2caiSeg: Semantic Segmentation of Laparoscopic Images using Convolutional Neural Networks | 2020 | Salman Maqbool | Cholecystectomy | Laparoscopic | Liver, gall bladder, fat, upper wall, intestine, artery |
| Deep Neural Network-Based Semantic Segmentation of Microvascular Decompression Images | 2021 | R Bai | Microvascular decompression surgeries focusing on cerebral vessel and cranial nerve | Microscopic | Trigeminal nerve, facial nerve, glossopharyngeal nerve, vagus nerve, anterior inferior cerebellar artery, posterior inferior cerebellar artery, petrosal vein |
| Object and anatomical feature recognition in surgical video images based on a convolutional neural network | 2021 | Y Bamba | Abdominal surgeries, including colorectomy, rectal surgery, hernia repair, and sigmoid resection. | Laparoscopic | Gastrointestinal tract, blood, vessels, and uterus. |
| Automated segmentation by deep learning of loose connective tissue fibers to define safe dissection planes in robot-assisted gastrectomy | 2021 | Y Kumazu | Robot-assisted gastrectomy with a focus on suprapancreatic lymph node dissections. | Robot | loose connective tissue fibers |
| Development of an artificial intelligence system using deep learning to indicate anatomical landmarks during laparoscopic cholecystectomy | 2021 | T Tokuyasu | Cholecystectomy | Laparoscopic | Cystic duct, common bile duct, lower edge of the left medial liver segment, and Rouviere's sulcus. |
| Video recognition of simple mastoidectomy using convolutional neural networks: Detection and segmentation of surgical tools and anatomical regions | 2021 | J Choi | Simple mastoidectomy | Open | External auditory canal, antrum, tegmen, sigmoid sinus, spine of Henle |
| Development of a Visual Guidance System for Laparoscopic Surgical Palpation using Computer Vision | 2021 | KG Caballas | Cholecystectomy | Laparoscopic | Gallbladder, uterus |
| Anatomical Landmarks Detection for Laparoscopic Surgery Based on Deep Learning Technology | 2021 | AA Pozdeev | Renal Surgery | Laparoscopic | Renal arteries |
| Cortical Vessel Segmentation for Neuronavigation Using Vesselness-Enforced Deep Neural Networks | 2021 | N Haouchine | Craniotomy for brain tumor resection | Microscopic | Cortical vessels |
| Computer-assisted real-time automatic prostate segmentation during TaTME: a single-center feasibility study | 2021 | Daichi Kitag uchi | Transanal total mesorectal excisions | Laparoscop ic | Prostate |
| Using deep learning to identify the recurrent laryngeal nerve during thyroidectomy | 2021 | Julia Gong | Thyroidectomy, with or without neck dissection | Open | Recurrent laryngeal nerve |
| Effective semantic segmentation in Cataract Surgery: What matters most? | 2021 | Theodoros Pissas | Cataract Surgery | Microscopi c | Pupil, iris, cornea, skin |
| Real-time vascular anatomical image navigation for laparoscopic surgery: experimental study | 2022 | D Kitaguchi | Sigmoid Colectomy or High Anterior Resection | Laparoscop ic | Inferior mesenteric artery, superior rectal artery |
| Artificial Intelligence-Based Total Mesorectal Excision Plane Navigation in Laparoscopic Colorectal Surgery | 2022 | T Igaki | Total Mesorectal Excision during Rectal resection | Laparoscop ic | Areolar tissue in the total mesorectal excision plane |
| Deep-Learning-Based Cerebral Artery Semantic Segmentation in Neurosurgical Operating Microscope Vision Us ing Indocyanine Green Fluorescence Videoangiography | 2022 | M Kim | NeurosurgicalProc edures | Microscopi c | Cerebral vessels (arteries or veins) |
| A deep learning framework for real-time 3D model registration in robot-assisted laparoscopic surgery | 2022 | E Padovan | Robot-Assisted Radical Prostatectomy and Robot-Assisted Partial Nephrectomy. | Robot | Prostate and kidney |
| Deep Learning Based Real-Time Semantic Segmentation of Cerebral Vessels and Cranial Nerves in Microvascular Decompression Scenes | 2022 | R Bai | Microvascular decompression procedures for trigeminal neuralgia and hemifacial spasm | Microscopi c | Trigeminal nerve, facial nerve, glossopharyngeal nerve, vagus nerve, anterior inferior cerebellar artery, posterior inferior cerebellar artery, petrosal vein |
| Analysis of Current Deep Learning Networks for Semantic Segmentation of Anatomical Structures in Laparoscopic Surgery | 2022 | B Silva | Cholecystectomy | Laparoscop ic | Liver, gallbladder, abdominal wall, gastrointestinal tract, fat, blood, connective tissue, hepatic vein, liver ligament, and cystic duct. |
| Concept Graph Neural Networks for Surgical Video Understanding | 2022 | Y Ban | Cholecystectomy | Laparoscop ic | Gall bladder, cystic artery, cystic duct, liver, cystic plate |
| Latent Graph Representations for Critical View of Safety Assessment | 2022 | A Murali | Cholecystectomy | Laparoscop ic | Gallbladder, cystic duct, cystic artery, hepatocystic triangle, cystic plate |
| Real-Time Detection of Recurrent Laryngeal Nerves Using Artificial Intelligence in Thoracoscopic Esophagectomy | 2022 | K Sato | Esophagectomy | Thoracosco pic | Recurrent laryngeal nerve |
| Space Squeeze Reasoning and Low-Rank Bilinear Feature Fusion for Surgical Image Segmentation | 2022 | ZL Ni | Cataract Surgery and Laparoscopic | Microscopi c and Laparoscop ic | Iris, pupil, cornea, skin. parenchyma, kidney, and intestinal in EndoVis2018. |
| Automatic Assessment of the Critical View of Safety in Laparoscopic Cholecystectomy Using Deep Learning | 2022 | Pietro Mascagni | Cholecystectomy | Laparoscop ic | Gallbladder, cystic duct, cystic artery, cystic plate, hepatocystic triangle |
| Deep Learning-Based Seminal Vesicle and Vas Deferens Recognition in the Posterior Approach of Robot-Assisted Radical Prostatectomy | 2022 | Nobushige Takeshita | Robot-assisted radical prostatectomy | Robot | Seminal vesicle and vas deferens |
| Deep learning-based recognition of key anatomical structures during robot-assisted minimally invasive esophagectomy | 2023 | RB den Boer | Esophagectomy | Robot | Azygos vein, vena cava, aorta, right lung |
| Fast instruments and tissues segmentation of micro-neurosurgical scene using high correlative non-local network | 2023 | YW Luo | Meningioma resection | Microscopi c | Blood Vessel, dura mater, and tumor area. |
| Artificial Intelligence in Minimally Invasive Adrenalectomy: Using Deep Learning to Identify the Left Adrenal Vein | 2023 | B Sergun | Transabdominal left adrenalectomy | Laparoscopic | Left adrenal vein |
| Anatomy segmentation in laparoscopic surgery: comparison of machine learning and human expertise - an experimental study | 2023 | FR Kolbinger | Anterior rectal resections and Rectal extirpations | Robot | Abdominal wall, colon, intestinal vessels (inferior mesenteric artery and inferior mesenteric vein with their subsidiary vessels), liver, pancreas, small intestine, spleen, stomach, ureter, and vesicular glands. |
| Deep-learning-based semantic segmentation of autonomic nerves from laparoscopic images of colorectal surgery: an experimental pilot study | 2023 | S Kojima | Left-sided colorectal resections for sigmoid colon and rectal cancers (Sigmoidectomy, LAR, HAR, APR). | Laparoscopic | Autonomic nerves (hypogastric nerve and superior hypogastric plexus) |
| Developing the surgeon-machine interface: using a novel instance-segmentation framework for intraoperative landmark labelling | 2023 | JJ Park | Spinal dural arteriovenous fistula surgery. | | Dorsal spinal arteries, spinal cord |
| Feature Tracking and Segmentation in Real Time via Deep Learning in Vitreoretinal Surgery: A Platform for Artificial Intelligence-Mediated Surgical Guidance | 2023 | RG Nespolo | Vitreoretinal surgeries like core vitrectomy, membrane peeling, and endolaser application. | Microscopic | Optic disc, fovea, retinal tears, retinal detachment, epiretinal membrane, fibrovascular proliferation, endolaser spots, and global area where endolaser was applied. |
| Towards automatic verification of the critical view of the myopectineal orifice with artificial intelligence | 2023 | M Takeuchi | Trans Abdominal Pre-Peritoneal inguinal hernia repair. | Laparoscopic | Pubic symphysis, direct hernia orifice, Cooper's ligament, iliac vein, triangle of Doom, deep inguinal ring, iliopsoas muscle |
| Detecting the location of lung cancer on thoracoscopic images using deep convolutional neural networks | 2023 | Y Ishikawa | Lung Resections | Thoracoscopic | Lung surfaces, tumor lesions on the pulmonary surface |
| Artificial intelligence for the recognition of key anatomical structures in laparoscopic colorectal surgery | 2023 | D Kitaguchi | Colorectal surgery | Laparoscopic | Ureter and autonomic nerve |
| Dynamic Scene Graph Representation for Surgical Video | 2023 | F Holm | Cataract surgery | Microscopic | Pupil, iris, cornea, skin |
| Automated Assessment of Critical View of Safety in Laparoscopic Cholecystectomy | 2023 | Y Li | Cholecystectomy | Laparoscopic | Liver, cystic artery, gall bladder, cystic plate, cystic duct, and fat. |
| An intraoperative artificial intelligence system identifying anatomical landmarks for laparoscopic cholecystectomy: a prospective clinical feasibility trial (J-SUMMIT-C-01) | 2023 | Hiroaki Nakanuma | Cholecystectomy | Laparoscopic | Cystic duct, common bile duct, Rouviere's sulcus, and the lower edge of the left medial hepatic segment (S4) |
| Towards reliable hepatocytic anatomy segmentation in laparoscopic cholecystectomy using U-Net with Auto-Encoder | 2023 | Koloud N. Alkhamaiseh | Cholecystectomy | Laparoscopic | Cystic duct, cystic artery, gallbladder, and liver |
| Deep learning-based vessel automatic recognition for laparoscopic right hemicolectomy | 2024 | K Ryu | Right Colectomy | Laparoscopic | Superior mesenteric vein, ileocolic artery, and ileocolic vein |
| Intraoperative artificial intelligence system identifying liver vessels in laparoscopic liver resection: a retrospective experimental study | 2024 | N Une | Liver resection | Laparoscopic | Hepatic veins and Glissonean pedicles |
| Deep Learning Model for Real-time Semantic Segmentation During Intraoperative Robotic Prostatectomy | 2024 | SG Park | Radical prostatectomy | Robot | Bladder, prostate, and seminal vesicle-vas deferens. |
| PitSurgRT: real-time localization of critical anatomical structures in endoscopic pituitary surgery | 2024 | Z Mao | Pituitary surgery. | Endoscopic | Sella, clival recess and carotid artery, and optic protuberance, |
| SurgNet: Self-Supervised Pretraining With Semantic Consistency for Vessel and Instrument Segmentation in Surgical Images | 2024 | J Chen | Hepato-pancreatic-biliary surgeries, (hepatectomy, pancreaticoduoden | Laparoscopic | Inferior pancreaticoduodenal artery, Henle trunk, proper hepatic artery, common hepatic artery, gastroduodenal artery, right gastroepiploic vein, portal vein, superior mesenteric vein |
| | | | ectomy, and resection of hilar cholangiocarcinom a | | |
| Real-time segmentation of biliary structure in pure laparoscopic donor hepatectomy | 2024 | N Oh | Pure laparoscopic donor hepatectomy | Laparoscop ic | Bile duct, transection site |
| Artificial intelligence for surgical safety during laparoscopic gastrectomy for gastric cancer: Indication of anatomical landmarks related to postoperative pancreatic fistula using deep learning | 2024 | Y Aoyama | Gastrectomy | Laparoscop ic | Pancreas, mesogastrium, transverse mesocolon, intestine, blood vessels, greater omentum, pre-pancreatic fatty tissue |
| Distinguishing the Uterine Artery, the Ureter, and Nerves in Laparoscopic Surgical Images Using Ensembles of Binary Semantic Segmentation Networks | 2024 | N Serban | Hysterectomy | Laparoscop ic | Uterine artery, the ureter, and nerves |
| Surgical optomics: hyperspectral imaging and deep learning towards precision intraoperative automatic tissue recognition—results from the EX-MACHYNA trial | 2024 | E Bannone | Laparotomies | Open | Hepatic ligament, mesentery, colon, bile duct, stomach, liver, gallbladder, pancreas, skin, artery, vein, omentum, and small bowel. |
| Intraoperative detection of parathyroid glands using artificial intelligence: optimizing medical image training with data augmentation methods | 2024 | JH Lee | Thyroidectomy (unilateral lobectomy). | Open | Parathyroid glands |
| Detection of Anatomical Landmarks During Laparoscopic Cholecystectomy Surgery Based on Improved Yolov7 Algorithm | 2024 | Z Yang | Cholecystectomy | Laparoscop ic | Gallbladder, Calot's triangle, and common bile duct |
| Interactive Surgical Training in Neuroendoscopy: Real-Time Anatomical Feature Localization Using Natural Language Expressions | 2024 | NM Matasyoh | Endoscopic third ventriculostomy | Endoscopic | Choroid plexus, foramen of Monroe, anterior septal vein, mamillary bodies, infundibular recess |
| An Artificial Intelligence-Based Nerve Recognition Model for Surgical Support and Education in Laparoscopic and Robot-Assisted Rectal Cancer Surgery | 2024 | Kazuya Kin oshita | Rectal cancer surgeries | Robot, Laparoscop ic | Nerve-like structures |
| Accuracy of Thoracic Nerves Recognition for Surgical Support System Using Artificial Intelligence | 2024 | Junji Ichinose | Lung Resections | Thoracosco pic | Thoracic nerves |
| SwinD-Net: a lightweight segmentation network for laparoscopic liver segmentation | 2024 | Shuiming Ouyang | Cholecystectomy | Laparoscop ic | Liver |

LAR, low anterior resection; HAR, high anterior resection; APR, abdominoperineal resection.

**Supplementary Table 2**: A description of data contained in all studies included in the review.

| Title | Year | SOTA model | Other models | Scores | GPU | Inference time per frame |
|---|---|---|---|---|---|---|
| CaDIS: Cataract dataset for surgical RGB-image segmentation | 2019 | HRNetV2. | UNet, DeepLabV3+, and UPerNet | **mIoU**: Pupil - 94.1, Iris - 84.6, Cornea - 93.2, Skin-70 | Two NVIDIA GTX 1080 Ti | |
| SurgAI: deep learning for computerized laparoscopic image understanding in gynaecology | 2020 | Mask R-CNN | | **mIoU**: Uterus - 84.5%, ovary - 29.6% | GPU - GTX 1080 | |
| NephCNN: A deep-learning framework for vessel segmentation in nephrectomy laparoscopic videos | 2020 | NephCNN, which is based on a 3D Fully Convolutional Neural Network (FCNN) architecture- 3D U-Net-inspired architecture with spatio-temporal feature extraction and adversarial training for shape-constrained segmentation | 2D U-net, 3D U-net | **Median DSC**: 71.76% (NephCNN), **Precision**: 94.8% (3D-Unet), **Recall**: 51.6% (NephCNN) | NVIDIA RTX 2080TI with 11GB memory | |
| Deep learning for semantic segmentation of organs and tissues in laparoscopic surgery | 2020 | TernausNet11 with VGG-11 encoder, and LinkNet with ResNet encoders | LinkNet, FCN-8s | **Mean IoU**: 78.3% | NVIDIA GeForce GTX 1070Ti | |
| m2caiSeg: Semantic Segmentation of Laparoscopic Images using Convolutional Neural Networks | 2020 | 10-layer CNN Encoder-Decoder | | **Range IoU**: 0.00-0.73, **Range Precision**: 0.01-0.78, **Range Recall**: 0.00-0.92, **Range F1**: 0.00-0.85 (on m2caiSeg)<br><br>**Range IoU**: 0.0-75.8, **Range Precision**: 0.0-88.6, **Range Recall**: 0.0-0.84, **Range F1**: 0.0-0.86 (On EndoVis 2018) | | |
| Deep Neural Network-Based Semantic Segmentation of Microvascular Decompression Images | 2021 | DeepLabv3+ with feature distillation block (FDB) and atrous spatial pyramid pooling (ASPP) (Xception-65 backbone with the FDB and ASPP module). | U-Net, PSPNet, DANet, and FastFCN | **mIOU** (over all classes): 75.73%, **Range IoU**: 62.1-87.8% | NVIDIA GEFORCE RTX 2080Ti | |
| Object and anatomical feature recog nition in surgical video images based on a convolutional neural network | 2021 | YOLO V3 | Faster R-CNN, tiny YOLO V2, Detectron, Single Shot Detector (SSD) and structured segment network (SSN) | For object detection, **Recall** and **Precision**: GI tract - 92.9 and 91.3; blood - 50 and 80; vessel -79.3 and 82.1; uterus - 75 and 90. | | |
| Anatomical Landmarks Detection for Laparoscopic Surgery Based on Deep Learning Technology | 2021 | YOLOv4-tiny and SSD object detection models | | **Mean average precision**: 0.62 | | |
| Cortical Vessel Segmentation for Neuronavigation Using Vesselness-Enforced Deep Neural Networks | 2021 | U-Net architecture for segmentation, enhanced with a vesselness-enforced network to improve vessel continuity | U-net-PB, Mask RCNN | **IoU**: 0.744 and **DSC** 0.85 | NVIDIA GeForce GTX 1070 | |
| Using deep learning to identify the recurrent laryngeal nerve during thyroidectomy | 2021 | Mask R-CNN | | **Mean DSC**: 0.71 | | |
| Effective semantic segmentation in Cataract Surgery: What matters most? | 2021 | OCRNet (Resnet50Encoder, OCRHead with Lovasz loss and RF sampling) | UPerNet, HRVNet, and DeepLabv3+ | **mean IoU**: 71.9% | NVIDIA Quadro P6000 with 24GB memory for UPN models, NVIDIA Quadro RTX 8000 GPU with 48GB memory for OCRNet | |
| Artificial Intelligence-Based Total Mesorectal Excision Plane Navigation in Laparoscopic Colorectal Surgery | 2022 | DeepLabv3+ | | **DSC**: 0.84 | | |
| Deep-Learning-Based Cerebral Artery Semantic Segmentat ion in Neurosurgical Operating Microscop e Vision Using Indocyanine Green Fluorescence Videoangiography | 2022 | DeepLabv3+ - Fully convolutional network (FCN)-base network created by changing the last linear layer of Visual Geometry Group (VGG) network into a 1×1 convolution layer and by adding upscaling to predict each pixel (backbone network, ResNet-101) | FCN, DeepLabv3 and Unet | **Mean DSC** 0.77, **MeanIoU**: 0.63, **Accuracy**: 0.76, **TP**: 9.99%, **FP**: 4.67%, **TN**: 76.1%, **FN**: 9.23% | NVIDIA GeForce GTX 1080 Ti with 11 GB of VRAM | |
| A deep learning framework for real-time 3D model registration in robot-assisted laparoscopic surgery | 2022 | UNet-ResNet for segmentation and a modified ResNet50 for rotation estimation. | | **Mean IoU**: 0.73 for prostate and 0.86 for Kidney. | NVIDIA Quadro P4000 | |
| Analysis of Current Deep Learning Networks fo r Semantic Segmentation of Anatom ical Structures in Laparoscopic Surgery | 2022 | U-Net++ | A mix of conventional CNN architectures (U-Net, DynUNet, DeepLabV3+) and transformer-based models (UNETR). | **DSC** 0.35-0.91. **MeanIoU**: 0.55, **Recall**: 0.72, **Precision**: 0.73 | Tesla V100 40GB | |
| Concept Graph Neural Networks for Surgical Video Understanding | 2022 | Concept Graph Neural Network (ConceptNet) incorporating ResNet50 and Vision Transformer (ViT) as visual backbones for feature extraction (with LSTM-based nodes and hyperedges for structured learning) | Tripnet, Attention Tripnet, Rendezvous, ConceptNet-Resnet50, MTL, Resnet50-LSTMI3D, CSN, TSM | Object detection – **Average Accuracy**: CVS - 67.1, cystic artery - 55.1%, cystic duct - 55.2%, liver - 60.2%, cystic plate - 55.6%. | | |
| Latent Graph Representations for Critical View of Safety Assessment | 2022 | Graph Neural Network (GNN) applied to latent graph representations generated from object detection models and encoder based on ResNet50 and DeepLabv3+. | DeepCVS, DeepCVS-MobileNetV3, DeepCVS-ResNet50 (Only CVS criteria | CVS Identification – **mAP**: 59.1% (DeepCVSResNet18), **mAP with segmentation masks**: 67.7% (LG-CVS). | NVIDIA V100 32GB | |
| | | | prediction), LayoutCVS, DeepCVS, ResNet50-DetInit (CVS identification with segmentation masks), and Mask-RCNN, Cascade Mask-RCCN, and DeepLabv3+ (Semantic Segmentation and object detection) | **DSC** 74.9% (Mask2Former) and detection **mean AP**: 34.8%. | | |
| Real-Time Detection of Recurrent Laryngeal Nerves Using Artificial Intelligence in Thoracoscopic Esophagectomy | 2022 | DeepLabv3+ | | **DSC** 0.58 | | |
| Automatic Assessment of the Critical View of Safety in Laparoscopic Cholecystectomy Using Deep Learning | 2022 | Two-stage deep learning model, consisting of a segmentation network DeepLabv3+ with an Xception 65 backbone, and a 6-layer neural network as the classification model. | | **Mean IoU**: for all tissue classes - 31.5-88.5%. **Mean AP** for CVS all criteria: 71.9% and **bACC**: 71.4%. **Sensitivity and specificity**: 70.4% and 72.4%. | | |
| Artificial Intelligence in Minimally Invasive Adrenalectomy: Using Deep Learning to Identify the Left Adrenal Vein | 2023 | Efficient Stage-wise Feature Pyramid Network (ESFPNet) with Mix Transformer (MiT) in an encoder-decoder structure. | | **Highest mean DSC**: 0.77 (±0.16 SD) with B2 model and 0.66 (±0.21) for B4 model. | | |
| Towards automatic verification of the critical view of the myopectineal orifice with artificial intelligence | 2023 | DETR (Detection Transformer) model for object detection. | | **Average Accuracy Range**: 55.3 - 94.7, **Sensitivity:** 58.8-100, **Average Precision Range**: 50-94.6% , **Specificity**: 33.3-90, | | |
| Detecting the location of lung cancer on thoracoscopic images using deep convolutional neural net works | 2023 | Scaled-YOLOv4, with a backbone, neck, and head architecture for extracting and processing multi-scale feature maps for object detection. | | **Precision**: 91.9%, **Recall**: 90.5%, **F1**: 91.1%, **AP**: 0.880 at a threshold value of 0.5. | Four NVIDIA Quadro RTX8000 | |
| Dynamic Scene Graph Representation for Surgical Video | 2023 | Mask2Former and OCRNet were used for segmentation, and a graph convolutional network (GCN) was applied for surgical phase recognition | DeepPhase for phase recognition | **mIoU**: 83.1% | Without GPU | |
| Automated Assessment of Critical View of Safety in Laparoscopic Cholecystectomy | 2023 | Transformer-based segmentation models, including a two-stream segmentation approach. | | **mIoU**: GB - 91.4%, Liver - 89.1%, cystic duct - 68.3%, cystic artery - 44.8 cystic plate - 57.1%. **Overall mIoU**: 74.5%, **Accuracy**: 93.7%, and **DSC** 84.3%.<br><br>For CVS criteria – **Accuracy**: 92%, **Balanced accuracy**: 54%, **PPV**: 23%, **NPV**: 94%. | | |
| Deep Learning Model for Real-time Semantic Segmentation During Intraoperative Robotic Prostatectomy | 2024 | Modified U-Net architecture with 6 added layers for improved feature extraction. | | **Mean DSC** bladder, prostate, and seminal vesicle–vas deferens- 0.74, 0.85, and 0.84. | | |
| SurgNet: Self-Supervised Pretraining With Semantic Consistency for Vessel and Instrument Segmentation in Surgical Images | 2024 | SurgNet model (UPerNet), a Transformer-based encoder-decoder architecture for pseudo-mask segmentation with a Pyramid Vision Transformer (PVT) as the encoder | MoCo V3, MoBY, SimMIM, UM-MAE (self-supervised methods), and Unet, Unet++, DnUNet, UNETR, Deeplabv3+ | **IoU**: 0.57 and **DSC** 0.72 | NVIDIA A100 | |
| Distinguishing the Uterine Artery, the Ureter, and Nerves in Laparoscopic Surgical Images Using Ensembles of Binary Semantic Segmentation Networks | 2024 | U-Net (segmentation models for each organ) with the EfficientNet-b3 encoder. | | **Jaccard**: 0.80-0.85, **DSC** 0.89-0.92 for individual organs. | NVIDIA Tesla K80 with 12 GB of memory | |
| Surgical optomics: hyperspectral imaging and deep learning towards precision intraoperative automatic tissue recognition—results from the EX-MACHYNA trial | 2024 | CNN | | **Median DSC** across all tissue classes: 50%, External validation step **median DSC** 50% (range, 24-93%). Validation step 2 - **median DSC** 94% in one center and 87% in another center (range, 33-92%). | | |
| Artificial Intelligence for Intraoperative Guidance: Using Semantic Segmentation to Identify Surgical Anatomy During Laparoscopic Cholecystectomy | 2020 | GoNoGoNet and CholeNet - ResNet50 followed by a multi-scale pyramid pooling module, which aggregates the feature maps from the CNN at four different scales (1×1, 2×2, 3×3 and 6×6) using the extracted frames. A Pyramid Scene Parsing Network (PSPNet) was used for pixel-wise semantic segmentation. | | **Range IoU**: 0.65-0.86, **Range F1/DSC** 0.70-0.92, **Range Accuracy**: 0.93-0.95, **Range Sensitivity**: 0.69-0.93, **Range Specificity** - 0.94-0.98 | NVIDIA 1080Ti | <10ms |
| Artificial intelligence for the recognition of key anatomical structures in laparoscopic colorectal surgery | 2023 | UreterNet (based on Feature Pyramid Networks and EfficientNetB7) and NerveNet (using Object-Contextual Representations Net and High-Resolution Net). | | **DSC**: 0.72 ureter, 0.58 hypogastric plexus, 0.63 - aortic plexus | Tesla T4 with 16 GB of VRAM | 100ms |
| Intraoperative detection of parathyroid glands using artificial intelligence: optimizing medical image training with data augmentation methods | 2024 | RetinaNet with ResNet backbone architecture for object detection and DeepFill v2 for image inpainting. | | **Average precision**: 0.79 and **F1 score**: 0.775. **Recall**: 0.85, **Precision**: 0.72. | 4 NVIDIA Tesla V100, each with 16 GB of memory. | 101ms |
| An Artificial Intelligence-Based Nerve Recognition Model for Surgical Support and Education in Laparoscopic and Robot-Assisted Rectal Cancer Surgery | 2024 | U-Net CNN | | **mean DSC** 0.44, **mean IoU**: 0.29 | Tesla V100 32GB | 16.7ms |
| Fast instruments and tissues segmentation of micro-neurosurgical scene using high correlative non-local network | 2023 | Encoder-decoder architecture with a two-branch design (detail and semantic branches) using a high correlative non-local network (HCNNet) with short-term dense concatenate (STDC) modules. | FastFCN, MobileNetv3-LRASPP, FastSCNN, ERFNet, MobileNetv2-FCN, ICNet, | For Tissues – **IoU Range**: 27.4-59.03% | GPU TITAN X Pascal | 18.2ms |
| | | | Cgnet, BiSeNetV2, HR18s-FCN, HR18s-OCR, STDCNet, BiSeNetV1 | | | |
| Automated segmentation by deep le arning of loose connective tissue fibers to define safe dissection planes in robot-assisted gastrectomy | 2021 | U-Net | | **Mean DSC** 0.549, range 0.335–0.691, **Recall**: 0.61, range 0.23-0.86 | Tesla V100 with 32 GB memory | 200ms |
| Artificial intelligence for surgical sa fety during laparoscopic gastrectomy for gastric cancer: Indication of anatomical landmarks related to postoperative pancreatic fistula using deep learning | 2024 | HyperSeg, a nested U-Net architecture for real-time segmentation. | | **DSC** 0.70 for pancreas across all scenes. | Four Tesla V100 | 210ms |
| Towards reliable hepatocytic anatomy segmentation in laparoscopic cholecystectomy using U-Net with Auto-Encoder | 2023 | U-Net integrated with an Auto-Encoder | | **Accuracy:** 95, **IoU**: 93, **Precision**: 97, **Mean IoU**; 80.9, **Hausdorff distance**: 15.8 | Tesla P100-PCIE-16 GB | 24ms |
| Feature Tracking and Segmentation in Real Time via Deep Learning in Vitreoretinal Surgery: A Platform for Artificial Intelligence-Mediated Surgical Guidance | 2023 | YOLACT++ with a ResNet-50 backbone for feature extraction, a feature pyramid network, a prediction head network, and a prototype generator network for segmentation. | | Object detection - **AUPR**: 0.68-0.97, Segmentation **AUPR**: 0.66-0.93. | NVIDIA RTX 2070Max-Q | 25.8ms |
| Deep learning-based recognition of key anatomical structures during robot-assisted minimally invasive esophagectomy | 2023 | CNN based on U-net architecture with an EfficientNet-B0 encoder. | | **DSC** 0.74-0.89, **Accuracy**: 0.97, **Sensitivity**: 0.85-0.95, **Specificity**: 0.98 | NVIDIA GTX Titan with 12 GB of internal memory | 26ms |
| Anatomy segmentation in laparosco pic surgery: comparison of machine learning and human expertise - an experimental study | 2023 | DeepLabv3 with ResNet50 backbone (Encoder-decoder structure) and SegFormer-based model for semantic segmentation. | | **IoU**: 0.31-0.85 (SegFormer), **F1**: 0.43-0.91, **Precision**: 0.40-0.90, **Recall**: 0.48-0.94, **Specificity**: 0.95-1.00 | NVIDIA A5000 | 28ms |
| Interactive Surgical Training in Neuroendocopy: Real-Time Anatomical Feature Localization Using Natural Language Expressions | 2024 | Combination of ResNet for image feature extraction and DistilBERT for language feature extraction, with a transformer-based fusion approach. | TransVg, SeqTR, QRNet, HCCAN | **Accuracy:** 93.67 and **mIoU**: 76.1% | NVIDIA GeForce RTX 4090 | 28ms |
| PitSurgRT: real-time localization of critical anatomical structures in endoscopic pituitary surgery | 2024 | PitSurgRT- HRNet-based architecture with multitask heads for segmentation and landmark detection. | PAINet, HRNetv2 | **IoU:** Sella - 67 and Clival Recess - 45.97. **Precision**: Sella - 81.3, Clival recess - 71.7. **Landmarks MPCK20**: 97.9 and **distance in pixels**: 54.5. | NVIDIA DGX A100 with 48 GB memory | 3.3ms |
| Accuracy of Thoracic Nerves Recognition for Surgical Support System Using Artificial Intelligence | 2024 | U-Net CNN | | **Dice index**: 0.56, **Jaccard index**: 0.39 | Tesla V100 with 32 GB memory | 33.3ms |
| Space Squeeze Reasoning and Low-Rank Bilinear Feature Fusion for Surgical Image Segmentation | 2022 | SRBNet - U-Net-like architecture with a dilated ResNet backbone, combined with the space squeeze reasoning module (SSRM) and low-rank bilinear | Unet, LinkNet, RAUNet, RefineNet, PAN, BARNet, PSPNet, DeepLabv3+ | **mDSC** 89.5 and **mIoU**: 82.42 (for Cataseg) **mDSC** 71.9 and **mIoU**: 62.2 (for EndoVis2018). | Tesla V100 | 38.5ms |
| | | fusion module (LBFM) for segmentation tasks | | For tissues only – **mDSC** 80.1-99.2 (RAUNet & NLBNet) and **mIoU**: 66.9-98.5 (RAUNet & NLBNet). | | |
| Deep Learning-Based Seminal Vesicle and Vas Deferens Recognition in the Posterior Approach of Robot-Assisted Radical Prostatectomy | 2022 | DeepLabV3+ | | **DSC**: 0.73, **Precision**: 0.74, **Recall**: 0.73, **FPR**: <0.01, **FNR**: 0.27 | NVIDIA Tesla V100S | 60ms |
| Real-time segmentation of biliary structure in pure laparoscopic donor hepatectomy | 2024 | DeepLabV3+ with ResNet50 encoder. | | **Mean DSC** - for bile duct 0.728 and Anterior wall for transection - 0.43, **Precision**: 0.71 for BD and 0.37 for AW, **Recall**: 0.74 for BD and 0.53 for AW. | NVIDIA GeForce RTXTM 3060 with 12GB of VRAM | 66.7ms |
| Deep Learning Based Real-Time Semantic Segmentation of Cerebral Vessels and Cranial Nerves in Microvascular Decompression Scenes | 2022 | MVDNet (with dilation), a custom-designed network with a U-shaped encoder-decoder architecture, using Light Asymmetric Bottleneck (LAB) and Feature Fusion Module (FFM). | Enet, ESPNet, FSSNet, CGNet, EDANet, ContextNet, DABNet | **Mean IoU**: 77.45% | NVIDIA 2080Ti | 7.3ms |
| Intraoperative artificial intelligence s ystem identifying liver vessels in laparoscopic liver resection: a retrospective experimental study | 2024 | Feature Pyramid Network with EfficientNetV2-L as the network backbone. | | **Mean DSC value:** 2-class model - 0.789 (vessels vs background). **Mean DSC**: hepatic vein and the Glissonean pedicle in the 3-class model - 0.631 and 0.482. | NVIDIA Tesla T4 with 16 GB of VRAM | 75ms |
| Deep-learning-based semantic segmentation of autonomic nerves from laparoscopic images of colorectal surgery: an experimental pilot study | 2023 | DeepLabV3+ with an Xception backbone. | | **DSC** 0.56 for HGN and 0.49 for SHP. **Precision**: 0.63 for HGN and 0.61 for SHP. **Recall**: 0.50 for HGN and 0.42 for SHP. **Average recognition rate:** 75.0%, which was calculated from 75.0% for the right HGN, 66.7% for the left HGN and 83.3% for the SHP. | NVIDIA Tesla V100 with 32 GB | 80ms |
| Deep learning-based vessel automatic recognition f or laparoscopic right hemicolectomy | 2024 | CNN | | **DSC** 0.54-0.78, **Precision**: 0.53-0.80, **Recall**: 0.55-0.77 | | 80ms |
| Detection of Anatomical Landmarks During Laparoscopic Cholecystectomy Surgery Based on Improved Yolov7 Algorithm | 2024 | Enhanced version of the YOLOv7 object detection algorithm with a channel attention mechanism. | YOLOv7, PP-PicoDet | **mAP**: 69.3% and **Precision**: 76.8%. | NVIDIA A30 | 83.3ms |
| Real-time vascular anatomical image navigation for laparoscopic surgery: experimental study | 2022 | DeepLabv3+ with ResNeSt-269 as the network backbone. | | **Mean DSC**: 0.798 | NVIDIA Quadro GP100 with 16 GB of VRAM | 83ms |
| Developing the surgeon-machine interface: using a novel instance-segmentation framework for intraoperative landmark labelling | 2023 | SOLOv2 + Resnet101DCN | MaskRCNN + Feature Pyramid Networks (FPN), SparseRCNN, YOLOv3 | **F1 score (Dice)**: 17% **mAP**: 15.2. | GeForce RTX 2080 Ti | 88ms |
| Computer-assisted real-time automatic prostate segmentation | 2021 | DeepLab v3+ | | **DSC** 0.71 for two colors and 0.676 for one | NVIDIA Quadro GP 100 | 90ms |
| during TaTME: a single-center feasibility study | | | | | | |
| An intraoperative artificial intelligence system identifying anatomical landmarks for laparoscopic cholecystectomy: a prospective clinical feasibility trial (J-SUMMIT-C-01) | 2023 | YOLOv3 | | **DSC**: CBD - 36%, cystic duct - 36%, median lobe of liver lower border - 49%, Rouvier's sulcus - 41%. | | 90ms |
| SwinD-Net: a lightweight segmentation network for laparoscopic liver segmentation | 2024 | CNN Encoder-decoder structure similar to U-Net, with added depthwise separable convolutions and Swin Transformer Blocks for capturing global features<br><br>Best model - DeepLabV3+ | SwinD-Net, U-Net, U-Net++, Attention U-net, TransUnet, MobileNetV3, UNeXt | **IoU**: 0.96,<br>**DSC** 0.98 | NVIDIA RTX 3090 of 24GB memory | 96ms |
| Development of an artificial intelligence system using deep learning to indicate anatomical landmarks during laparoscopic cholecystectom y | 2021 | YOLOv3. | | **Average Precision**: common bile duct: 0.320, cystic duct: 0.074, lower edge of the left medial liver segment: 0.314, and Rouviere's sulcus: 0.101. | Tesla V100 40GB | Designed for real-time, average speed of 37.2 frames per second. |
| Development of a Visual Guidance System for Laparoscopic Surgical Palpation using Computer Vision | 2021 | YOLACT++ | YOLO9000, Mask-RCNN | **Average Precision** (for Mask): 88.4% | NVIDIA GTX 2060 6GB VRAM | Offline. 20.6fps. |
| Video recognition of simple mastoidectomy using convolutional neural networks: Dete ction and segmentation of surgical to ols and anatomical regions | 2021 | YOLACT | YOLOv4, U-Net | Simultaneous object detection and segmentation for anatomical structures – **Range mAP**: 0.27-0.53 for anatomical structures (YOLACT).<br>**Mean DSC** 0.40 (YOLACT).<br><br>Segmentation only –<br>**Range DSC**: 0.28-0.80 (U-Net) and 0.26-0.64 (YOLACT). | TITAN RTX | YOLAC T - 32.3 frames per second (FPS). YOLOv4 - 58.5fps. U-Net 141.5fps. |

bACC, Balanced accuracy; DSC, dice score coefficient; FNR, false negativity rate; FPR, false positivity rate; mAP, mean average precision; mIoU, mean intersection over union; NPV, negative predictive value; PPV, positive predictive value; SOTA, state of the art.

**Supplementary Table 3**: The segmentation metrics (dice similarity coefficient and mean intersection-over-union (mIoU)) of all the organs evaluated in the studies included in the review.

| **Organs** | **Median Dice similarity coefficient (n-studies targeting the organ for segmentation)** | **Mean Intersection-over-Union (mIoU) (n-studies targeting the organ for segmentation)** |
|---|---|---|
| | | |
| Gall Bladder | 0.66 (0.57-0.84) (4) | 0.88 (0.70-0.91) (3) |
| Liver | 0.88 (0.82-0.98) (6) | 0.81 (0.75-0.89) (2) |
| Cystic duct | **0.49** (1) | 0.63 (0.58-0.68) (2) |
| Cystic artery | | **0.43** (0.43-0.44) (2) |
| Cystic plate | | **0.44** (0.31-0.57) (2) |
| Hepatocystic triangle | 0.79 (1) | |
| Superior mesenteric vein | 0.78 (1) | |
| Ileocolic artery | 0.55 (1) | |
| Ileocolic vein | 0.54 (1) | |
| Azygous vein | 0.79 (1) | |
| Aorta | 0.74 (1) | |
| Right lung | 0.89 (1) | |
| Inferior mesenteric artery | 0.69 (0.60-0.79) (2) | |
| Hepatic veins | 0.63 (1) | |
| Glissonean pedicles (1) | **0.48** (1) | |
| Cranial nerve 5 | | 0.82 (0.81-0.84) (2) |
| Cranial nerve 7 | | 0.82 (0.78-0.87) (2) |
| Cranial nerve 9 | | 0.78 (0.77-0.79) (2) |
| Cranial nerve 10 | | 0.81 (0.81-0.82) (2) |
| Anterior inferior cerebellar artery | | 0.71 (0.71-0.71) (2) |
| Posterior inferior cerebellar artery | | 0.74 (0.74-0.74) (2) |
| Petrosal vein | | 0.63 (0.62-0.64) (2) |
| Areolar tissue/fat | 0.84 (0.71-0.91) (3) | 0.77 (1) |
| Cerebral vessels | 0.82 (0.79-0.85) (2) | **0.35** (1) |
| Dura mater | | 0.59 (1) |
| Uterus | | 0.84 (1) |
| Ovary | | **0.29** (1) |
| Left adrenal vein | 0.77 (1) | |
| Pupil | | 0.94 (1) |
| Iris | | 0.84 (1) |
| Cornea | | 0.93 (1) |
| Prostate | 0.78 (0.71-0.85) (2) | 0.73 (1) |
| Kidney | 0.86 (0.86-0.86) (1) | 0.86 (1) |
| Abdominal wall | 0.84 (0.57-0.91) (3) | |
| Colon | 0.57 (0.49-0.79) (2) | |
| Intestinal veins | **0.49** (0.33-0.65) (2) | |
| Pancreas | 0.52 (0.47-0.70) (3) | |
| Small intestine | 0.71 (0.12-0.89) (3) | |
| Spleen | 0.85 (1) | |
| Stomach | 0.59 (0.43-0.75) (2) | |
| Ureter | 0.72 (0.58-0.72) (3) | |
| Vesicular glands | **0.43** (1) | |
| Bladder | 0.74 (1) | |
| Seminal vesicle-vas deferens | 0.84 (1) | |
| Hypogastric nerves | 0.57 (0.56-0.58) (2) | |
| Aortic plexus | 0.63 (1) | |
| Superior hypogastric plexus | **0.49** (1) | |
| Sella | | 0.67 (1) |
| Clival recess | | **0.46** (1) |
| Loose connective tissue | 0.55 (1) | |
| Vessels | 0.72 (1) | |
| Bile duct/Common bile duct | 0.72 (0.24-0.73) (3) | |
| External auditory canal | **0.46** (1) | |
| Antrum | **0.46** (1) | |
| Tegmen | **0.29** (1) | |
| Sigmoid sinus | 0.58 (1) | |
| Spine of Henle | 0.80 (1) | |
| Uterine artery | 0.92 (1) | |
| Nerves | 0.56 (0.44-0.56) (3) | |
| Hepatic ligament | 0.53 (1) | |
| Mesentery | **0.44** (1) | |
| Artery | 0.51 (1) | |
| Vein | 0.70 (1) | |
| Omentum | **0.18** (1) | |
| Recurrent laryngeal nerve | 0.64 (0.58-0.71) (2) | |
| Renal artery | 0.71 (1) | |